\documentclass[12pt]{elsarticle}
\usepackage{graphicx}
\usepackage{dcolumn}
\usepackage{bm}
\usepackage{subfigure}  
\usepackage{ctable}

\IfFileExists{srcltx.sty}{\usepackage[active]{srcltx}}

\newcommand{\be}{\begin{equation}}
\newcommand{\ee}{\end{equation}}
\newcommand{\bea}{\begin{eqnarray}}
\newcommand{\eea}{\end{eqnarray}}

\begin{document}


\title{
On the use of X-ray and $\gamma$-ray telescopes for identifying the origin of electrons
and positrons observed by ATIC, Fermi, and PAMELA}

\author[ucla]{Antoine~Calvez}
\author[ucla]{Warren~Essey}
\author[kc]{Malcolm~Fairbairn}
\author[ucla,ipmu]{Alexander~Kusenko}
\author[umd,nasa]{Michael~Loewenstein}

\address[ucla]{Department of Physics and Astronomy, University of California, Los
Angeles, CA 90095-1547, USA}
\address[kc]{Physics Department, King's College, London, UK}
\address[ipmu]{IPMU,
University of Tokyo, Kashiwa, Chiba 277-8568, Japan}
\address[umd]{Department of Astronomy,University of Maryland, College
Park, MD, USA}
\address[nasa]{CRESST and X-ray Astrophysics Laboratory NASA/GSFC, Greenbelt, MD, USA}


\begin{abstract}
X-ray and $\gamma$-ray observations can help understand the origin of the
electron and positron signals reported by ATIC, PAMELA, PPB-BETS, and Fermi. It remains unclear
whether the observed high-energy electrons and positrons are produced by relic
particles, or by some astrophysical sources. To distinguish between
the two possibilities, one can compare the electron population in the local
neighborhood with that in the dwarf spheroidal galaxies, which are not expected to
host as many pulsars and other astrophysical sources. This can be accomplished using X-ray and $\gamma$-ray
observations of dwarf spheroidal galaxies. Assuming the signal detected by Fermi and ATIC comes from
dark matter and using the inferred dark matter profile of the Draco dwarf
spheroidal galaxy as an example, we calculate the photon spectrum produced by electrons via inverse Compton scattering. Since little is known about the
magnetic fields in dwarf spheroidal galaxies, we consider the propagation of charged particles with and without diffusion.
Extending the analysis of Fermi collaboration for Draco, we find that for a halo mass $\sim10^{9}\,\mathrm{M}_{\odot}$, even in the absence of diffusion, the $\gamma$-ray signal would be above the upper limits. This conclusion is subject to uncertainties associated with the halo mass.  
If dwarf spheroidal galaxies host local magnetic fields, the diffusion of the electrons can result in a signal detectable by future X-ray telescopes.
\end{abstract}

\maketitle

\section{Introduction}

The nature of cosmological dark matter remains a tantalizing puzzle~\cite{Bertone:2004pz}.
If dark matter is made up of weakly interacting massive particles (WIMPs),
their annihilation products may be observed and used for identification of the dark-matter particles.
PAMELA~\cite{PAMELA}, ATIC~\cite{ATIC}, PPB-BETS~\cite{Torii:2008xu}, and Fermi~\cite{Abdo:2009zk}
have observed unexpected features in the electron and positron spectra at high energies.
The high-energy electrons and positrons could come from the annihilations or decays of dark matter
particles~\cite{Ibe:2008ye,Chen:2008md,Chen:2008qs,Fox:2008kb,ArkaniHamed:2008qn,Shirai:2009wi,Ibe:2009dx,Ibe:2009en,Shirai:2009wi,Shirai:2009kh}, but they
could also be produced by astrophysical sources, such as pulsars, supernova remnants and Gamma-Ray Bursts (GRBs)
\cite{Yuksel:2008rf,Ioka:2008cv,Profumo:2008ms,Stawarz:2009ig,Calvez:2010fd,PhysRevD.80.123017,PhysRevLett.103.111302,PhysRevLett.103.051104}.
Theoretical models of dark matter can accommodate a wide range of parameters
(see for example Refs.~\cite{ModelInd,FermiInt,Profumo:2008ms}). Astrophysical models of particle acceleration by pulsars are also
uncertain, but they can also account for the observed  signal. To
distinguish between the two possibilities, it would be desirable to compare the
electron and positron populations in the local neighborhood with that in some
other parts of the galaxy, which are known to be devoid of pulsars and other
potential astrophysical sources of high-energy particles. Dwarf Spheroidal
Galaxies (dSphs) present such an opportunity, provided that one can infer the photon spectra generated by
the interaction of electrons and positrons with the Cosmic Microwave Background radiation (CMB).

Observations of dwarf spheroidal galaxies have recently been used in a dedicated search for {\em decaying} dark matter
in the form of sterile neutrinos~\cite{Loewenstein:2008yi,Loewenstein:2009cm}. Sterile neutrinos are expected to undergo a
two-body decay, producing a narrow line in the X-ray spectrum (for a recent review, see, e.g., Ref.~\cite{Kusenko:2009up}).
Detection of WIMP, which are much heavier and which annihilate rather than decay, presents a very different challenge~\cite{Feng:2010gw}.
For a number of WIMP models, the X-ray and $\gamma$-ray signals would be too faint to observe in the foreseeable future, but the same models would
predict the flux of high-energy electrons and positrons well below the levels observed by ATIC, PAMELA, PPB-BETS, and Fermi.
A Breit-Wigner resonance~\cite{Ibe:2008ye} or long-range interactions~\cite{ArkaniHamed:2008qn,Lattanzi:2008qa} could increase the
dark matter annihilation cross section, but it is difficult to reconcile Sommerfeld enhancement with the primordial relic abundance of
WIMP or with the $\mu$-type distortion of the CMB energy spectrum~\cite{Feng:2009hw,Zavala:2009mi}. One does expect a boost factor from the
small scale structure of dark matter, but the required values are well in excess of one's expectations based on numerical N-body simulations.

In the absence of a compelling theoretical framework, we will not try to relate our predictions to any specific model of dark matter,
but we will focus on a model-independent determination of whether the high-energy electrons originate from dark matter
(which is abundant in both the local neighborhood and in a  dwarf spheroidal galaxy), or from some astrophysical source candidates
(whose population in a dwarf spheroidal galaxy is suppressed).

\section{Photon spectra}

Let us consider the spectrum of CMB photons up-scattered to the X-ray and $\gamma$-ray bands by the Inverse
Compton Scattering (ICS) of electrons and positrons, produced by the annihilation of dark-matter particles, in dwarf spheroidal galaxies.
For any self-annihilating dark matter candidate, we define the number density per unit time and energy of a certain
species $i$ in the final state of the reaction as
\begin{equation}
Q_{i}(\vec{r},E)=\left\langle\sigma
v\right\rangle_{0}\frac{\rho^{2}_{\chi}(\vec{r})}{2m^{2}_{\chi}}N_{i}(E),
\label{func:Q}
\end{equation}
$N_{i}(E)$ describes the energy dependent differential number of particles of species $i$ produced during each annihilation process. $N_{i}(E)$, the mass of the dark matter particle $m_{\chi}$, and the annihilation rate at zero temperature $\left\langle\sigma v\right\rangle_{0}$ are the only parameters in Equation (\ref{func:Q}) that are sensitive to the dark matter model. Note however that in reality, $\left\langle\sigma v\right\rangle_{0}$ is strongly constrained by the dark-matter abundance inferred from WMAP data~\cite{WMAP}.

The remaining parameter is the radius-dependent dark matter halo density profile $\rho_{\chi}(\vec{r})$, which we will take to be the Navarro, Frenk and White (NFW) profile~\cite{NFW} under the assumption that the dark matter is \emph{cold}, as is the case for WIMP candidates. In this study in particular, we will  consider the dark matter profile of the Draco dwarf Spheroidal~\cite{Mashchenko:2005bj} as an illustrative example of our technique and for comparison with previous work~\cite{DMXrayDwarf,MultiWaveDraco,Abdo:2010ex}. The Draco dSph is located at a distance $D_{\mathrm{dSph}}\sim 80\,\mathrm{kpc}$ from the earth. Based on~\cite{Mashchenko:2005bj} we will take the scaling density and the scaling radius of the NFW profile to be $\log\left(\frac{\rho_{s}}{\mathrm{M}_{\odot}\,\mathrm{kpc}^{-3}}\right)=7.20$ and $\log\left(\frac{r_{s}}{\mathrm{kpc}}\right)=0.45$. The corresponding virial mass of the dSph is $M_{\mathrm{vir}}=4.6\times10^{9}\,\mathrm{M}_{\odot}$.

To calculate the photon spectrum, we consider the production of electrons and positrons and their propagation in the interstellar medium until the point
where they interact. In theory, the process could be simulated using a numerical program such as GALPROP \cite{1998ApJ...509..212S}, but little is known about
the properties of the interstellar medium surrounding dwarf spheroidal galaxies. In light of these considerations, we will address
two different cases: with and without diffusion. Provided that the faintness of dSphs is due to their low stellar and gaseous content
\cite{Bowen:1996ga,Mateo:1998wg,Gallagher:2003nx}, the magnetic fields populating dwarf spheroidals are likely to be extremely small if at all existent.
Magnetic fields play a crucial role in the diffusion process, trapping charged particles in the local neighborhood of the galaxy and increasing their
interaction time. In the absence of magnetic fields the electrons and positrons created during the dark matter annihilation can escape freely.
We will therefore calculate the expected signal first for the case where the electrons interact with the CMB as they freely leave the dSph.
Second, for completeness, we will address the possibility that, although unlikely, magnetic fields do exist in the local neighborhood of dSphs;
we will thus also derive the expected signal under the assumption that the electrons and positrons are subject to diffusion.

\subsection{Spectrum in the absence of diffusion}

In the absence of magnetic fields, the electrons and positrons created during the annihilation process are free to escape the local neighborhood of the
dwarf spheroidal. These relativistic electrons have enough energy to up-scatter CMB photons to the X-ray and $\gamma$-ray bands through inverse Compton scattering. In the Klein-Nishina limit, where in the rest frame of the electron the energy of the photon is $\epsilon \gg m_{e}c^{2}$, the energy spectrum of
the up-scattered photon is \cite{PhotonSpec}:
\begin{eqnarray}
\frac{dN_{p}}{dt\,d\epsilon_{u}}(\epsilon_{u},\epsilon,\gamma_{e})=&&\frac{2\pi r_{0}^{2}c}{\gamma^2_{e}}\frac{n_{\mathrm{CMB}}(\epsilon)}{\epsilon}\\
&\times&\left[2q\ln q+(1+2q)(1-q)+\frac{\Gamma^{2}q^{2}}{2(1+\Gamma q)}(1-q)\right],\label{func:KN}\nonumber
\end{eqnarray}
%
where
\begin{equation}
\Gamma=\frac{4\epsilon\gamma_{e}}{m_{e}c^{2}},\qquad
q=\frac{E_{u}}{\Gamma(1-E_{u})},\qquad
E_{u}=\frac{\epsilon_{u}}{\gamma_{e}m_{e}c^{2}},\nonumber
\end{equation}
$\epsilon_{u}$ is the energy of the up-scattered photon, $n_{\mathrm{CMB}}(\epsilon)$ is the black-body spectrum of the CMB, and $r_{0}$ is the classical electron radius. The final energy of a photon going through inverse Compton scattering should not only depend on the initial energies of the photon and electron, but also on the angle of the initial collision and on the final scattering angle. In the case of the CMB, the photon bath in which the electron is traveling is isotropic, thus Equation (\ref{func:KN}) was obtained by averaging over all incoming photons angles, and by integrating over the total scattering solid angle.

As they escape, the electrons will lose energy through various processes (synchrotron, ICS, bremsstrahlung, ionization).
At the scale of the excess observed by Fermi and PAMELA, the dominant source of loss is ICS, and the rate at which the electrons dissipate energy is given by:
\begin{equation}
b_{\rm IC}(E)=\left(0.25\times10^{-16}\,\textrm{GeV s}^{-1}\right)\left(\frac{E}{1\textrm{GeV}}\right)^{2}.\nonumber
\end{equation}
This is a really small effect. The typical distance between a dSph of the local group and the earth is on the order of $\sim100\,\textrm{kpc}$, and as mentioned previously, in our case, the distance to Draco is $D_{\mathrm{dSph}}\sim 80\,\mathrm{kpc}$. On that length scale, a $1\,\textrm{TeV}$ electron will not lose more then $10\%$ of its original energy. Therefore, although the ICS process is important from a physical point of view as it will give rise to the X-ray and $\gamma$-ray signals, it will only have a minimal effect on the kinetic energy of the escaping electron. From a mathematical standpoint, the no diffusion case is comparable to making $D(E)$ arbitrarily large and setting $b(E)$ to zero in Equation (\ref{func:DiffEQ}) (see Section \ref{sec:diffcase} for a description of the transport equation). To a good approximation, the photon flux can be obtained by making the unphysical assumption that the electrons go through the inverse Compton process without being subjected to any energy losses. We determine the flux at earth by calculating the flux going through the surface of a sphere of radius $D_{\mathrm{dSph}}$ centered on the Draco dSph. The interaction time of each electron as it moves radially outward to the surface of the sphere with constant energy is $t_{int}\sim\frac{D_{\mathrm{dSph}}}{c}$. The flux at earth is then:
\begin{equation}
\mathcal{F}_{\rm{ND}}(\epsilon_{u})\sim\frac{1}{4\pi D_{\rm{dSph}}^{2}}\int
dE\,d\epsilon\,dV\,Q_{e}(\vec{r},E)\frac{dN_{p}}{dt\,d\epsilon_{u}}\left(\frac{D_{\rm{dSph}}}{c}\right)\nonumber,
\label{func:FluxND}
\end{equation}
The volume integral is performed over the halo of dSph up to a distance of 2.5~kpc, and the two integrals over the incoming electron and photon
energies are performed over the kinematically allowed range~\cite{PhotonSpec}.

\subsection{Spectrum in the presence of diffusion \label{sec:diffcase}}

To determine the differential electron density in the presence of magnetic fields we will follow
Refs.~\cite{DMXrayDwarf,FreqAnalysis,MultiWaveDraco} and model the
diffusion and energy loss with the following transport equation:
\begin{eqnarray}
\frac{\partial}{\partial
t}\frac{dn_{e}}{dE}=&&\nabla\left[D(E,\vec{r})\nabla\frac{dn_{e}}{dE}\right]
+\frac{\partial}{\partial
E}\left[b(E,\vec{r})\frac{dn_{e}}{dE}\right]\nonumber\\
&&+Q_{e}(E,\vec{r}),\label{func:DiffEQ}
\end{eqnarray}
where $D(E,\vec{r})$ is the diffusion coefficient, $b(E,\vec{r})$ is
the energy loss term, and $\frac{dn_{e}}{dE}$ is the sum of the
differential electron and positron densities. Under the simplifying assumptions that the system has reached equilibrium, and that the
diffusion and the energy loss terms are spatially homogeneous,
Eq.~(\ref{func:DiffEQ}) becomes:
\begin{equation}
D(E)\nabla^{2}\frac{dn_{e}}{dE}+\frac{\partial}{\partial
E}\left[b(E)\frac{dn_{e}}{dE}\right]+Q_{e}(E,\vec{r})=0.\label{func:DiffEQ2}
\end{equation}
We will assume a diffusion term of the form~\cite{FreqAnalysis,Colafrancesco:1998us,Blasi:1999aj}:
\begin{equation}
D(E)=\frac{d_B^{2/3}}{B_{\mu}^{1/3}}D_0\left(\frac{E}{1\textrm{GeV}}\right)^{
\gamma},
\label{func:diffusion}
\end{equation}
where $d_B$ is the minimum scale of uniformity of the magnetic field, $B_{\mu}$
is the size of the magnetic field in $\mu$G and $D_0$ is a constant.

A slightly different model was considered in Ref.~\cite{1959SvA.....3...22S} to study the energy spectrum of synchrotron radiation in the Milky Way.
Although the study assumed a departure from equilibrium, and a constant diffusion coefficient, it found an electron spectrum similar to our results.

Based on the analysis of cosmic ray fluxes in the Milky Way Ref.~\cite{Diffusion} found  $0\leq\gamma\leq1$ with a preferred
value of $\gamma=0.7$. The value of $d_{B}^{2/3}D_{0}$ is not well known.
In the same reference, a median value
of $D_{0}=1.1\times10^{27}\textrm{cm}^2 \, \textrm{s}^{-1}$ was used. However, this parameter
depends on the magnitude and size of magnetic field
inhomogeneities, which are unknown for systems such as dwarf spheroidal
galaxies. Our choice of parameter will be discussed further is Section
\ref{sec:diff}.

The energy loss term is \cite{DMXrayDwarf,FreqAnalysis}
\begin{eqnarray}
b(E)=&&b_{\rm IC}^{0}\left(\frac{E}{1\textrm{GeV}}\right)^{2}\nonumber\\
&&+b_{\rm syn}^{0}\left(\frac{B}{1\mu\textrm{G}}\right)^{2}\left(\frac{E}{1\textrm{GeV}}
\right)^{2}\nonumber\\
&&+b_{\rm coul}^{0}n\left(1+\frac{1}{75}\log\left(\frac{\gamma_{e}}{n}
\right)\right)\nonumber\\
&&+b_{\rm brem}^{0}n\left[\log\left(\frac{\gamma_{e}}{n}\right)+0.36\right]\left(\frac{E}{1
\textrm{GeV}}\right),\nonumber
\end{eqnarray}
where $b_{\rm IC}^{0}=0.25\times10^{-16}\,\textrm{GeV s}^{-1}$, $b_{\rm syn}^{0}=0.0254\times10^{-16}\,\textrm{GeV s}^{-1}$,
$b_{\rm coul}^{0}=6.13\times10^{-16}\,\textrm{GeV s}^{-1}$ and
$b_{\rm brem}^{0}=1.51\times10^{-16}\,\textrm{GeV s}^{-1}$; $n$ defines
the electron thermal density and is taken to be
$n=10^{-6}\textrm{cm}^{-3}$, while $\gamma_{e}$ is the usual
relativistic $\gamma$-factor of the electron.

Eq.~(\ref{func:DiffEQ2}) can be solved exactly
\cite{FreqAnalysis}. The solution takes the form:

\begin{equation}
\frac{dn_{e}}{dE}(E,\vec{r})=\frac{1}{b(E)}\int_{E}^{M_{\chi}}dE'\,G(r,\Delta
v)Q_{e}(E',\vec{r}),\label{func:dndE}
\end{equation}
where
\begin{eqnarray}
G(r,\Delta v)&=&\frac{1}{\sqrt{4\pi\Delta
v}}\sum_{n=-\infty}^{\infty}(-1)^{n}\int_{0}^{r_{h}}dr'\,\frac{r'}{r_{n}}\frac{
\rho^{2}_{\chi}(r')}{\rho^{2}_{\chi}(r)}\label{func:Green}\\
&&\times\left[\exp\left(-\frac{(r'-r_{n})^{2}}{4\Delta
v}\right)-\exp\left(-\frac{(r'+r_{n})^{2}}{4\Delta
v}\right)\right].\nonumber
\end{eqnarray}
Here $r_{n}=(-1)^{n}r+2nr_{h}$, and $r_{h}$ is the diffusion radius. We will take
our diffusion radius to be twice the visible radius. Finally we will define $\Delta v=v(E)-v(E')$ with:
\begin{eqnarray}
v(\psi)=\int_{\psi_{\rm min}}^{\psi}dy\,D(y)\label{func:DiffLength}\\
\psi(E)=\int_{E}^{E_{\rm max}}\frac{d E'}{b(E')}\nonumber.
\end{eqnarray}
For the relevant parameter values for this problem, Eq.~(\ref{func:Green}) is safely convergent after $10$ iterations.
One finally obtains the overall photon spectrum by folding the differential electron density with the Klein-Nishina spectrum.
Again, the integrals have to be performed over the volume of the dSph, and for the allowed kinematic range:
\begin{equation}
\frac{dN_{X-ray}}{dt\,d\epsilon_{u}}(\epsilon_{u})=\int
dE\,d\epsilon\,dV\,\frac{dN_{p}}{dt\,d_{\epsilon_{u}}}\frac{dn_{e}}{dE}.
\end{equation}

The flux near the earth can then easily be calculated:
\begin{equation}
\mathcal{F}_{\rm D}(\epsilon_{u})=\frac{1}{4\pi
D_{D}^{2}}\frac{dN_{\rm X-ray}}{dt\,d\epsilon_{u}}(\epsilon_{u})\nonumber,
\label{func:FluxD}
\end{equation}

\section{Application}
\subsection{Source function from Pamela and Fermi}

Based on the Pamela and Fermi results, we would like to deduce a model independent source function, $Q(E,\vec{r})$, for the electrons created during the annihilation of the dark matter particles. If one assumes that the function doesn't change with time, then from Equation (\ref{func:Q}) one can write:
\begin{equation}
Q(E,\vec{r})=q(E)\rho^2_\chi(\vec{r}),
\end{equation}
where $q(E)$ contains all model dependent terms, including the boost factors.

If we consider the flux per unit energy caused by an infinitesimal element of
$Q$ and assume spherical symmetry we find,
\begin{equation}
\frac{d\mathcal{F}}{dE\,dr}=q(E)\rho^2_\chi(r)
\end{equation}
It seems reasonable to integrate this out to the typical propagation length of
the electrons given by \cite{FreqAnalysis}
\begin{equation}
l_{\rm prop}\approx\sqrt{\frac{D(E)E}{b(E)}},
\end{equation}
where the functions $D(E)$ and $b(E)$ were defined previously. We take $D_0=1.1\times
10^{27}\textrm{cm}^2 \textrm{s}^{-1}$, $\gamma=0.7$ and use IC scattering as the
dominant part of $b(E)$. Since $l_{\rm prop}$ is only a few $\textrm{kpc}$'s we take the local Milky Way value of
$\rho_\chi(r)$ as roughly constant at $0.35\,\textrm{GeV} \textrm{cm}^{-3}$ and
compare the total flux to those measured by Pamela and Fermi to get
\begin{equation}
q(E)=\beta \left(\frac{E}{1\,\textrm{GeV}}\right)^{-a}
\end{equation}
Where $\beta\sim 6\times
10^{-26}\textrm{GeV}^{-3} \textrm{cm}^{3} \textrm{s}^{-1}$ and $a\sim 1.88$.

\subsection{Choice of diffusion constant\label{sec:diff}}

Very little is known about the interstellar medium and magnetic fields in dwarf spheroidal galaxies.
The dynamo mechanism driven by differential rotation and the wind produced in supernovae are almost certainly
not operational in a dwarf spheroidal galaxy~\cite{1992ApJ...401..137P,2004ApJ...605L..33H}.
While it is possible that the dSphs could create their own galactic wind due to the energy output by stars~\cite{1987Natur.329..613B},
they should be strongly affected by ram pressure and stripping due to the wind of the mother galaxy~\cite{2008ApJ...674..258E}.
It is probably unlikely that appreciable magnetic fields could build up in dwarf spheroidal galaxies, and so the effects of diffusion should not be
very important. However, for the purpose of completeness we include diffusion in our analysis.
Since the dynamo action is almost certainly not scalable from the mother galaxy to dwarf galaxies, the value of $D_0$ can differ significantly
from what one obtains using the scaling arguments. If one assumes that the spectrum of fluctuations of the magnetic field follows a Kolmogorov spectrum,
one can show that the diffusion constant must be given by Eq.~(\ref{func:diffusion}). Since very
little is known about the magnetic field structure of dwarf spheroidal galaxies
we look at two different possibilities. First we will consider the preferred Milky value of $10^{27}\textrm{cm}^2 \textrm{s}^{-1}$ \cite{Diffusion}.
Furthermore, since lower magnetic fields generate higher diffusion coefficients, and because we expect the magnitude of the magnetic fields present
in the dSph to be very small we consider $10^{29}\textrm{cm}^2 \textrm{s}^{-1}$ as well. We also vary the magnetic field from $0.1-1.0\,\mu\textrm{G}$.

\subsection{Results}

We should emphasize that similar studies have been performed by the Fermi collaboration and by other groups \cite{DMXrayDwarf,MultiWaveDraco,Abdo:2010ex}, however the studies were never performed in a model independent manner and only included the diffusion dependent case, which we believe to be physically inaccurate.

\begin{figure}
\begin{center}
\includegraphics[width=95mm]{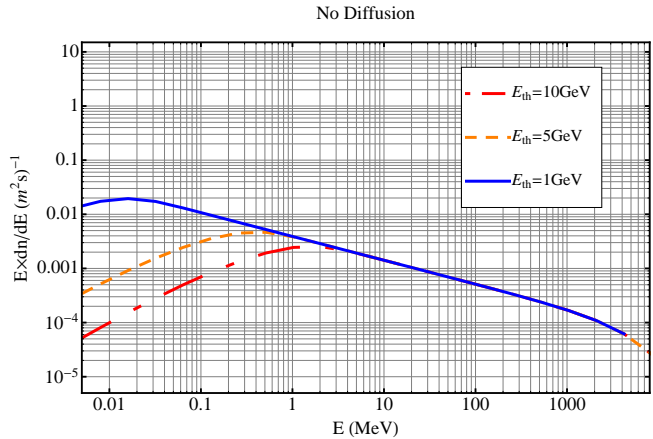}
\caption{\label{fig:ND} Photon spectra in the absence of diffusion, for $\gamma=0.3$.}
\end{center}
\end{figure}

\begin{figure}
\begin{center}
\includegraphics[width=95mm]{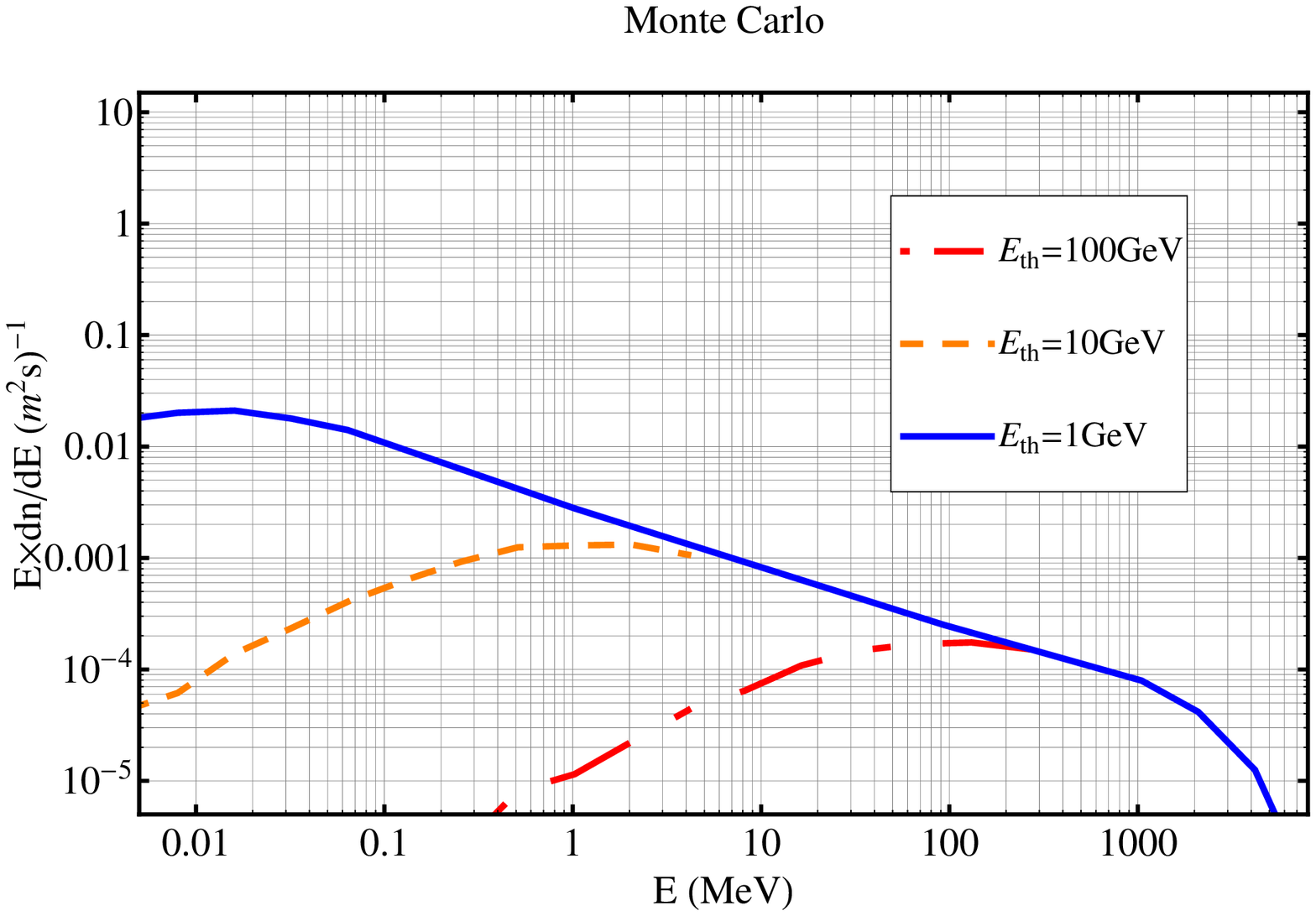}
\caption{\label{fig:MonND} Monte Carlo results for photon spectra in the absence of diffusion for a variety of low energy cutoffs.}
\end{center}
\end{figure}

\begin{figure}
\begin{center}
\begin{tabular}{cc}
\subfigure[\label{fig:FD27B1muG}]{
\includegraphics[width=70mm]{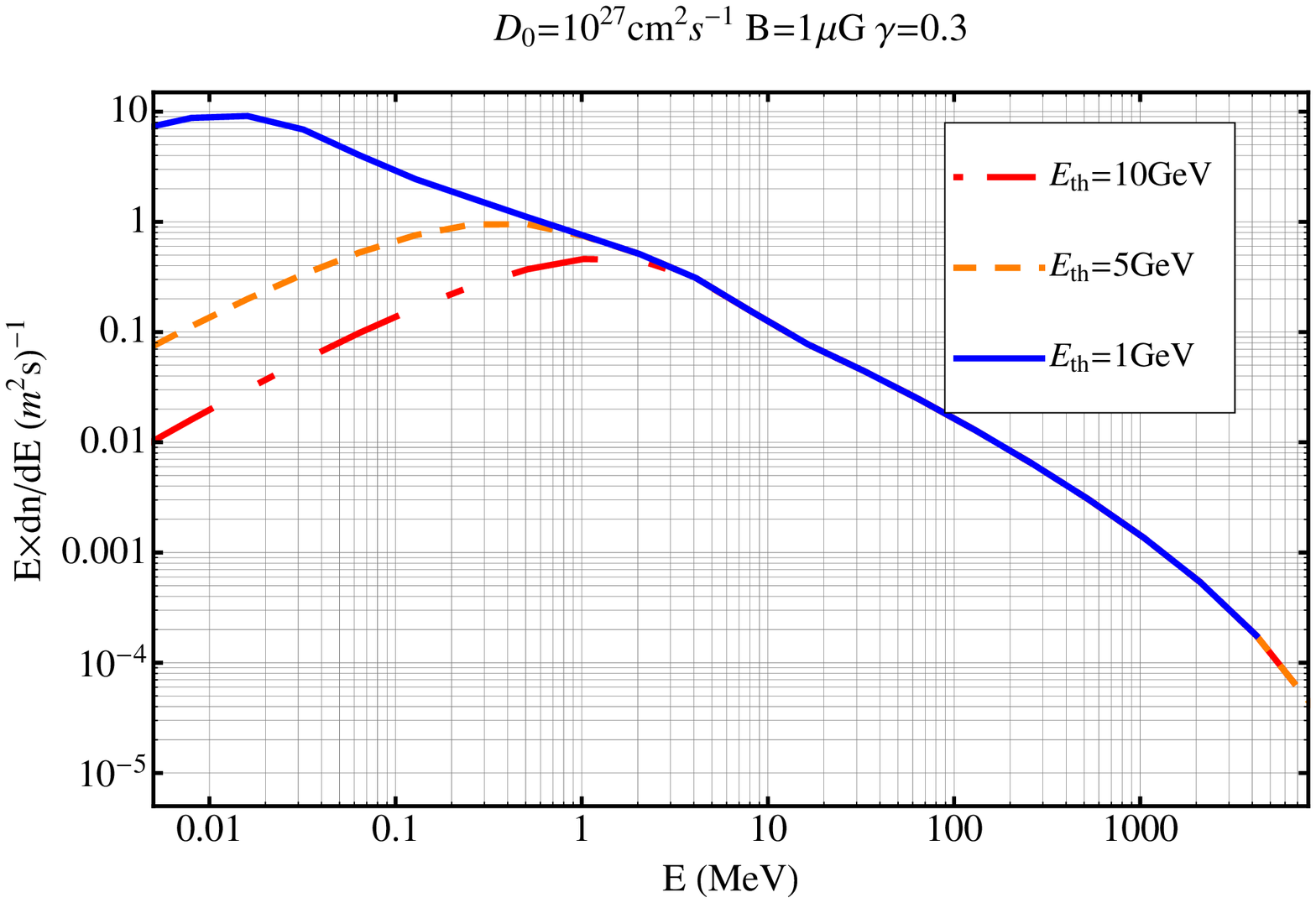}}
&
\subfigure[\label{fig:FD27B100nG}]{
\includegraphics[width=70mm]{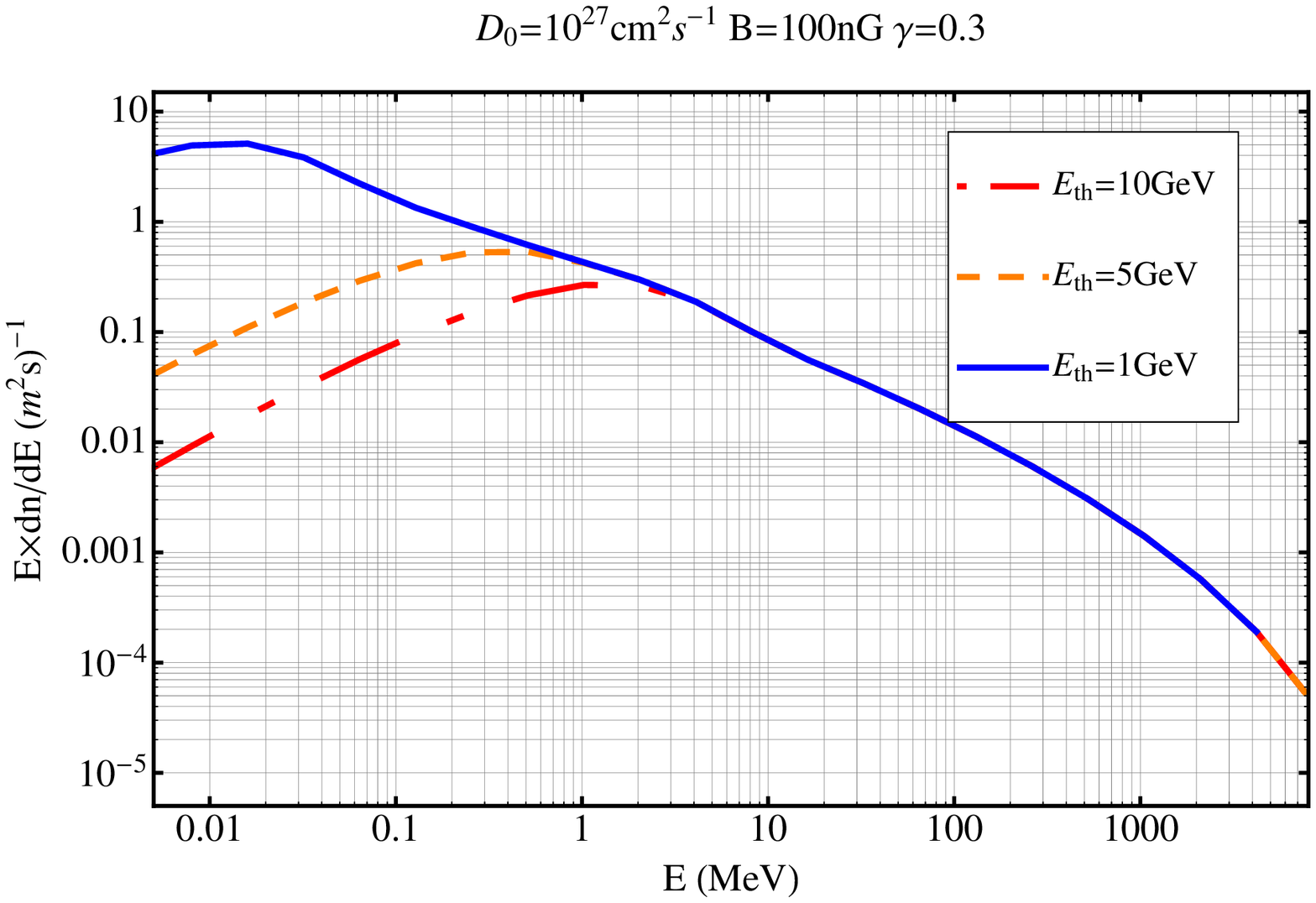}}
\\
\subfigure[\label{fig:FD29B1muG}]{
\includegraphics[width=70mm]{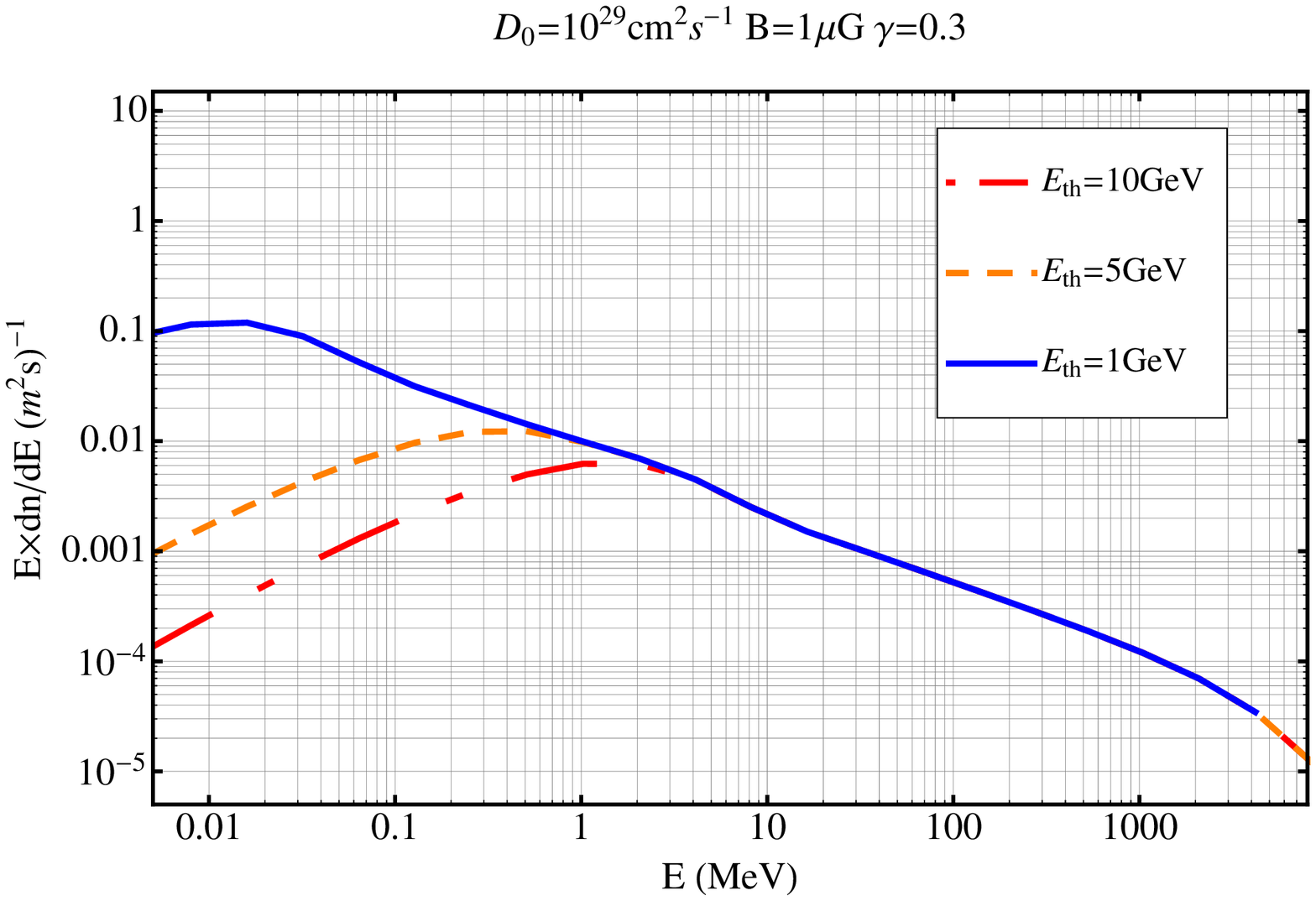}}
&
\subfigure[\label{fig:FD29B100nG}]{
\includegraphics[width=70mm]{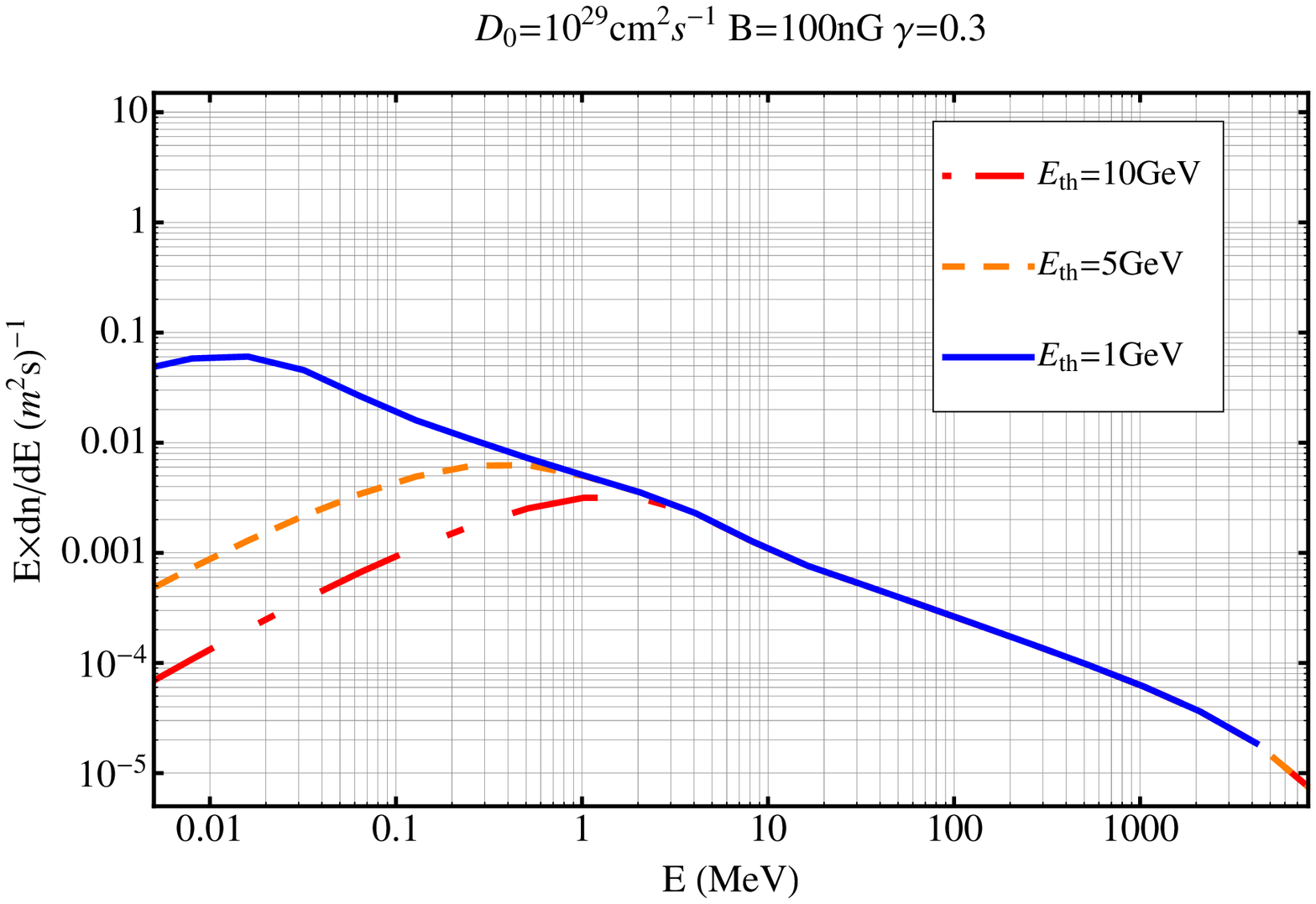}}
\end{tabular}
\caption{Photon spectra for $\gamma=0.3$ and for a low energy threshold of $1\,\mathrm{GeV}$ (blue, solid), $5\,\mathrm{GeV}$ (orange, dash), and $10\,\mathrm{GeV}$ (red, dot-dash). Figure \ref{fig:FD27B1muG}: diffusion constant $D_{0}=10^{27}\textrm{cm}^{2} \textrm{s}^{-1}$,  magnetic field $B=1\,\mu\textrm{G}$. Figure \ref{fig:FD27B100nG}: $D_{0}=10^{27}\textrm{cm}^{2} \textrm{s}^{-1}$, $B=100\,\textrm{nG}$. Figure \ref{fig:FD29B1muG}: $D_{0}=10^{29}\textrm{cm}^{2} \textrm{s}^{-1}$, $B=1\,\mu\textrm{G}$. Figure \ref{fig:FD29B100nG}:
$D_{0}=10^{27}\textrm{cm}^{2} \textrm{s}^{-1}$,  $B=100\,\textrm{nG}$.
}
\end{center}
\end{figure}

In Fig. \ref{fig:ND} through Fig. \ref{fig:FD29B100nG} we present the results of our analysis. Fig. \ref{fig:ND} presents the photon spectra in the absence of diffusion, while Fig. \ref{fig:FD27B1muG} through Fig. \ref{fig:FD29B100nG} present the cases where diffusion is present. Note that for those cases where we include diffusion, we only present results for a Kolmogorov spectral index $\gamma=0.3$. The spectra resulting from the case of the Milky Way preferred value, namely $\gamma=0.7$, though slightly lower in magnitude, were not significantly different from the ones presented here and were therefore omitted. As a systematic check, in the case of diffusionless transport, we calculated the photon spectra using a Monte Carlo which followed individual particle as they traveled from Draco to the earth. The result of the Monte Carlo simulation is presented in Fig. \ref{fig:MonND}.

Our results are very sensitive to the lower energy cutoff of the initial electron and positron populations (i.e. to the lower limit of the incoming electron energy integral in Eq. (\ref{func:FluxND}) and Eq. (\ref{func:FluxD})), thus we present our results for various \emph{threshold} energy values. Based on the shape of the excess presented by Fermi, we place an artificial cutoff on the incoming electron spectrum at $E=900\,\mathrm{GeV}$. The location of the high energy cutoff however, has very little effect on the resulting photon flux. This is in part due to the fact that the incoming photon flux is very low at high energies.

The predicted spectra in the case where diffusion is present is on
average significantly higher than in the absence of diffusion. This is because, as aforementioned, in the presence of diffusion the electrons
are effectively trapped by the magnetic fields, increasing their interaction time and consequently increasing the resulting photon
signal. As it can be seen from Fig. \ref{fig:FD27B1muG} through Fig. \ref{fig:FD29B100nG}, our predicted signal in the hard X-ray band is highly sensitive to the magnetic field strength, and the the electron positron signal low energy threshold is required to extend below $10\,\textrm{GeV}$. Simulations of the
predicted spectra as observed by the {\it Suzaku} Hard X-ray Detector (HXD) demonstrate that the dark matter signal, even under optimistic
assumptions, is overwhelmed by cosmic and internal backgrounds. The prospects are considerably improved with respect to the next
generation of hard X-ray telescopes. We have constructed simulated {\it Astro-H} spectra, adopting the optimal parameters for detection
(solid line in Fig. \ref{fig:FD27B1muG}), to assess the feasibility of targeting the Draco dwarf spheroidal, or a similar system, with this observatory.

With regard to the {\it Astro-H} Hard X-ray Imaging System \cite{2010SPIE.7732E..33K}, we consider the ``all-layers configuration'' $20-70\,\textrm{keV}$ band, and use simulated spectra to predict the source counts from Draco, as well as the cosmic X-ray and detector non-X-ray (particle) backgrounds (CXB
and NXB, respectively). The currently available spectral response and simulation files enable one to consider sources with X-ray emission
that is either spatially unresolved by the $\sim 1.5'$ half-power diameter mirror \cite{2010SPIE.7732E..32K} ($1.8'$ circular aperture spectral extraction
region assumed) or flat over (at least) four square degrees (full $8.6'\times 8.6'$ field-of-view extraction region assumed). In the
latter case, the predicted source count rate in the $20-70\,\textrm{keV}$ bandpass is $0.00135\,\textrm{ct}\,\textrm{s}^{-1}$, as compared to $\sim 0.0034\,\textrm{ct}\,\textrm{s}^{-1}$ from the cosmic X-ray background (CXB) and $\sim 0.0185\,\textrm{ct}\,\textrm{s}^{-1}$ for the detector non-X-ray (particle) background (NXB) -- so that deep exposures and highly accurate background calibration would be required. However, unlike the backgrounds that are flat, the dark
matter signal is highly peaked with $3-10\%$ of the signal enclosed within the central $0.03$ deg \cite{2011arXiv1104.0412C}. As a result, based on scaling using
the reduction in beam, and the vignetting derived from the provided effective area functions, we find that the signal-to-background ratio
may be boosted by as much as an order of magnitude with a judicious choice of spectral extraction region. The Nuclear Spectroscopic
Telescope Array ({\it NuStar}) mission has similar capabilities in terms of bandpass, sensitivity, background, and angular resolution to
the of the {\it Astro-H} HXI. The {\it Astro-H} Soft Gamma-ray Detector (SGD) \cite{2010SPIE.7732E..34T} is more sensitive still, despite its lack of
optics, due to the flat predicted source photon spectra, and observed $40\,\textrm{keV}$ cutoff of the CXB. Using the same assumptions as above, the predicted SGD Draco count rate is $0.012\,\textrm{ct}\,\textrm{s}^{-1}$ in the $40-300\,\textrm{keV}$ band. $100000$ sec SGD simulations predict that the flux and high energy
photon index may be determined to $\sim 10\%$ accuracies. In the event of a detection with Fermi, these hard X-ray and soft $\gamma$-ray observations could be used to constrain the propagation of the dark matter annihilation products.

The $\gamma$-ray signal should be observable by the Fermi satellite. The analysis done by the Fermi collaboration \cite{Abdo:2010ex}
sets flux upper-limits for the gamma-ray signal from various dwarf spheroidals. Assuming the $\gamma$-rays originating from the dSphs followed a power law:
\begin{equation}
\mathcal{F}_{\mathrm{dSph}}=\mathcal{F}_{0}\left(\frac{E}{E_{0}}\right)^{-\zeta},
\label{func:pwerlaw}
\end{equation}
and accounting for all other known sources in the field of view of the telescope at the time of the observation as well as for the galactic
diffuse emission and for the corresponding isotropic component, the Fermi collaboration performed a likelihood analysis to fit the observed spectrum. The power law presented in Eq. (\ref{func:pwerlaw}) is constrained by the overall normalization $\mathcal{F}_{0}$, and the spectral index $\zeta$, while $E_{0}$ is just an arbitrary energy scale. Their analysis revealed that their current data was consistent with $\mathcal{F}_{0}=0$, regardless of the value of the spectral index $\zeta$, for all observed dSphs. Despite the lack of clear signal from dwarf spheroidal galaxies, an upper limit on the flux was derived using a profile likelihood technique. The results of the analysis for the case of Draco, as they are presented in \cite{Abdo:2010ex} are shown in Table \ref{tab:Fermiflux}.
\begin{table}[htbp]
  \centering
    \begin{tabular}{ccccccccccc}
    \addlinespace
    \toprule
    Energy & \multicolumn{5}{c}{$E>100\,\mathrm{MeV}$} & \multicolumn{5}{c}{$E>1\,\mathrm{GeV}$} \\
    \midrule
    $\zeta$ & 1.0   & 1.8   & 2.0   & 2.2   & 2.4   & 1.0   & 1.8   & 2.0   & 2.2   & 2.4 \\
    $\mathcal{F}_{0}$  & 0.09  & 0.59  & 0.94  & 1.41  & 1.94  & 0.06  & 0.13  & 0.16  & 0.21  & 0.26 \\
    \bottomrule
    \end{tabular}%
    \caption{Upper-limit on the flux $\mathcal{F}_{0}$, in units of $10^{-5}\,\mathrm{m}^{-2}\,\mathrm{s}^{-1}$, coming from the draco dwarf spheroidal galaxy at the $95\%$ Confidence Level} \label{tab:Fermiflux}%
\end{table}%

The no-diffusion cases, Fig. \ref{fig:ND} and Fig. \ref{fig:MonND}, produce the lowest flux and are still comfortably above these upper limits. The $\textrm{GeV}$ signal is insensitive to the low energy cutoff of the original electron positron signal and is dominated by the IC scattering off the CMB. Thus, it becomes very difficult to reconcile these flux upper limits with the predicted signal without having a dark matter component in dwarf spheroidals that differs significantly from the local dark matter population. This could be a hint that astrophysical sources may be at the origin of the signal. However, the discrepancies could also be the result of the large uncertainty in the overall mass of the dSph, which would directly impact the NFW profile of the halo. 

\section{Conclusion}

Current and future observations of dwarf spheroidal galaxies with $\gamma$-ray detectors and with the next generation X-ray telescopes may be able
to probe the origin of high-energy electrons and positrons observed by Fermi and PAMELA. Current results from Fermi, which we have extended to the
possible case of non-diffusive propagation in a dwarf spheroidal galaxy, seem to be in conflict with the
most conservative $\gamma$-ray prediction from an almost model independent analysis of an assumed dark matter signal.
The litmus test for dark matter annihilations as the possible origin of these signals comes from comparing the high-energy particle fluxes in a
dwarf spheroidal galaxy and in the local neighborhood of the Milky Way disk.

Dark matter dominated systems, such as dwarf spheroidal galaxies, should generate a predictable flux of the dark-matter annihilations products, subject to mass model uncertainties. At the same time, most of the astrophysical sources capable of producing high-energy particles in the sun's neighborhood are absent in dwarf spheroidal galaxies, which have very few stars and very little gas as compared with their dark matter content. The high-energy electrons and positrons produced in dwarf spheroidal galaxies can generate X-rays and $\gamma$-rays by up-scattering CMB photons to higher energy bands. The predicted fluxes are above the upper limits set by Fermi for Draco; although this result could be interpreted as favoring an astrophysical source for the excess of high energy electrons and positrons in our galaxy, the large uncertainty in the mass of the halo limit the strength of this interpretation. Better mass measurements need to be obtained before conclusions can be made. 

\section{Acknowledgments}

We thank the Astro-H team for providing simulation tools. A.K. thanks P.~Biermann for helpful comments.  The work of A.C. and A.K. was supported in part by DOE grant DE-FG03-91ER40662 and by the NASA ATP grant NNX08AL48G. M.L. was supported by NASA ADAP grant \newline
NNX11AD36G. M.F. acknowledges support from the EU Marie Curie Network UniverseNet (HPRN-CT-2006-035863).

\bibliography{Pamela_Results}
\bibliographystyle{utphys}
\end{document}